\documentclass[a4paper,pdflatex]{article}
\usepackage[english]{babel}
\usepackage{graphicx,amsmath,amsfonts}

\begin{document}

\begin{center}

{\large \bf A QUASICLASSICAL APPROXIMATION IN THE THEORY OF THE
LANDAU--POMERANCHUK EFFECT }

\vspace{.25cm}

{\bf  A.V. Tarasov, H.T. Torosyan, and O.O. Voskresenskaya}

\vspace{.25cm}

{\small \it Joint Institute for Nuclear Researches, 141980 Dubna, Moscow region,
Russia
}

\end{center}


\begin{abstract}

\noindent Using the improved value of the screening angular
parameter in the quasiclassical approximation of the Moli\`{e}re
multiple scattering theory we show that the best agreement between
the Migdal theory of the LPM-effect and experiment is achieved if
the multiple scattering of electrons by target atoms is described
using the quasiclassical approximation instead of the traditionally
used Born one.

\end{abstract}


\bigskip

{\it  1. Introduction.} --- In 1953 Landau and Pomeranchuk
\cite{C-2} predicted within classical electrodynamics that multiple
scattering can considerably suppress bremsstrahlung of high energy
charged particles in a medium. The effect due to the fact that
multiple particle-atom scattering in a medium leads to destructive
interference of photons emitted at different space points, provided
that the formation (coherence) length $l_{cr}$ of the bremsstrahlung
is small in comparison with the mean free path of the particle in
the medium ($l_{cr}\ll L$, where $L$ is the thickness of a target).
This result was obtained on the basis of the classical theory of
electron radiation. However, the formula for a spectral density of
radiation in substance, obtained by Landau and Pomeranchuk, has only
evaluative character.

The quantitative theory of the multiple scattering effect on an
electron radiation in an amorphous medium was offered by Migdal in
\cite{C-4,C-4.5}. This theory was based on the application of the
kinetic equation method to the given task. Owing to Migdal's
important contribution to the theory of the given effect now it is
called the Landau--Pomeranchuk--Migdal  (LPM) effect.\footnote{Note
that Goldman has taken into account also the edge effects
\cite{C-5}.}

The further development of a quantitative LPM effect theory was
achieved  in \cite{Operat1} with use of the quasiclassical operator
method in QCD \cite{Operat2}. One of the basic equations of this
approach is a two dimensional Schr\"{o}dinger equation in the impact
parameter space with an imaginary potential obtained with use of
kinetic equations describing a motion of electron in a medium in the
presence of external field. The same equation (without external
field) was rederived in \cite{C-3}. The last derivation is based on
the approach \cite{Z1} results of which coincide basically with
\cite{Operat2}. This equation can be solved using a transverse Green
function based on a path integral. The criticism of the
quasiclassical operator approach \cite{Crit} was denied in
\cite{Oproverg}.

In \cite{Sh1} it was shown that analogous effects are possible also
at coherent radiation of relativistic electrons and positrons in a
crystalline medium, and the theory of these effects must also be
based on the quasiclassical methods. The LPM effect is relevant in
many areas of physics but particularly in high energy cosmic ray air
showers \cite{showers}. It has analogues in nuclear physics
involving quarks and gluons moving through matter \cite{KST}. A
LPM-type suppression also appears in stellar interiors \cite{stars}.

The first tests of LPM suppression came shortly after Migdal's paper
appeared; these previous experiments have studied the LPM effect,
mostly with cosmic rays \cite{prev}. They qualitatively confirmed
the LPM effect, but with very limited statistics. The first
quantitative measurement of the LPM effect for high energy electrons
was performed at SLAC in a series of experiments \cite{C-1.1}.
These experiments were the challenge for the theory since in all the
previous papers calculations
are performed to logarithmic accuracy;
but they are faced with an unexpected problem, the so-called
`problem of normali\-zation'.
The experimental data
obtained for 25 GeV electron beam interacting with a homogeneous
gold target disagree with the theoretical predictions within the
normalization factor 0.93~ \cite{C-3} -- 0.94~\cite{C-1.1}, and the
reasons of this disagreement are not clear.



However, considering the fact that the calculations \cite{C-3,C-1.1}
were performed within the Born approximation, the above-mentioned
discrepancy can be explained at least qualitatively. The aim o this
work is to shown that this discrepancy
can be explained also quantitatively if the corrections to the
results of the Born approximation are appropriately considered. In
this work it is shown that the use of a revised Moli\`{e}re theory
of multiple scattering \cite{C-11} within the quasiclassical
approach to the description of the particle-atom scattering allows
overcoming the discrepancies \cite{C-3,C-1.1} between experiment and
the Migdal LPM effect theory. Some results of this work are
presented in \cite{preprint}.

The paper is organized as follows: Section 2 provides the essential
background for our results. In this section, we get the basic
formulae of the LPM effect theory for finite-size targets in the
small-angle approximation adhering to the classical works by Migdal
and Goldman. The expressions for the intensity $dI(\omega)/d\omega$
of the bremsstrahlung produced the electron moving in matter and the
kinetic equation for the electron distribution function are given
within this theory in the small-angle approximation.
In Section 3 on the basis of the Moli\`{e}re theory and the
Fokker--Planck approximation, the analytical solution of the
equation for $dI(\omega)/d\omega$ at the emitted photon frequency
$\omega\gg\omega_{cr}$ is obtained.
This solution and the exact relation  between the values of the
screening angle $\theta_a$ of the Moli\`{e}re theory in the Born and
quasiclassical approximations are used to calculate
$(dI(\omega)/d\omega)_{qel}$ and $(dI(\omega)/d\omega)_{Born}$ and
estimate their ratio $R(\omega,L)$ for $\omega\gg\omega_{cr}$.
It is shown that the $R(\omega,L)$ value coincides with the
normalization constant $R$ from \cite{C-3,C-1.1}; however, the
latter ignores the dependence of the ratio on $\omega$ and $L$.
In Section 4 brief conclusions are given.

\bigskip


{\it 2. Formalism of the LPM effect theory for finite-size targets.}
--- Simple but cumbersome calculations based on the results of
\cite{C-4,C-5} yield the following formula for the electron
bremsstrahlung intensity averaged over various trajectories of
electron motion in the medium (hereafter the units $\hbar = c =1$,
$e^2=1/137$ are used):

$$\frac{dI}{d\omega} = 2\sum_{\vec \epsilon} \biggl\{ n_0 L
\int f^{\ast}(\vec n_2)\nu(\vec n_2-\vec n_1)f(\vec n_1) d\vec n_1 d\vec
n_2  $$
$$-(n_0v)^2\int\limits_{0}^{T}dt_1\int\limits_{t_1}^{T}dt_2 \,\, Re\biggl [
\int f^{\ast}(\vec n_2)\nu(\vec n_2-\vec n_2^{\prime})\,
w(t_2,t_1,\vec n_2^{\prime},\vec n_1^{\prime},\vec k)$$
\begin{equation}
\times f(\vec n_1)\nu(\vec n_1^{\prime}-\vec n_1)
d\vec n_1d\vec n_1^{\prime}d\vec n_{2}d\vec n_{2}^{\prime}\biggl
]\biggl\}~,
\end{equation}
where  $$f(\vec n_{1,2})=\frac{e}{2\pi}\cdot\frac {\vec \epsilon\vec
v_{1,2}}{1-\vec n \cdot \vec v_{1,2}}~,$$ $$\vec v_{1,2}=v\cdot \vec
n_{1,2}~,~~~\vec n = \frac{\vec k}{\omega}~,~~~ d\vec n_{1,2} \equiv
do_{1,2}~,~~~T=\frac{L}{v}~,$$ $$\nu(\vec n_2-\vec n_1)=\delta(\vec
n_2-\vec n_1) \int \sigma_0(\vec n_2^{\prime}-\vec n_1)d\vec
n_2^{\prime} -\sigma_0(\vec n_2-\vec n_1)~,$$
$$w(t_2,t_1,\vec n_2,\vec n_1,\vec k)= \int \tilde w(t_2,t_1,\vec r_2-\vec
r_1, \vec n_2, \vec n_1)$$ $$\times \exp[i\omega(t_2-t_1)-i\vec
k(\vec r_2-\vec r_1)]d\vec r_2~.$$
Here $\vec k$ and $\vec \epsilon$ are the wave vector and the
polarization vector of the emitted photon,
 $\vec n_{1,2}$ are the unit vectors in the electron motion direction, $v$ is the
electron velocity assumed to be invariant during the interaction
with the target (the quantum-mechanical recoil effect is negligibly
small), $e$ is the electron charge, $\sigma _0(\vec n_2 - \vec n_1)~
=~ d\sigma /do_{\vec n_2}$  is the differential cross-section of the
electron scattering  by target atoms, $n_0$ is the number of atoms
in an unit volume of the medium, $L $ is the thickness of the
target, $w(t_2,t_1,r_2-r_1, \vec n_2, \vec n_1)$ is the electron
distribution function  in the coordinates $\vec r_2$, and the
direction of motion $\vec n_2$ at the time $t_2$ provided that at
time $t_1$ the electron had the  coordinate $\vec r_1$ and moved in
the direction characterized by the unit vector $\vec n_1$.

The distribution function $w$ satisfies the kinetic equation
$$\frac{\partial w(t_2,t_1,\vec r_2-\vec r_1,\vec n_2,\vec
n_1)}{\partial t_2}=-\vec v_2 \cdot \vec \nabla_{\vec r_2}\cdot
w(t_2,t_1,\vec r_2-\vec r_1,\vec n_2,\vec n_1)$$
\begin{equation} -
n_0\int\nu(\vec n_2-\vec n_1^{\prime}) \tilde w(t_2,t_1,\vec
r_2-\vec r_1,\vec n_2^{\prime},\vec n_1)d\vec n_2^{\prime}
\end{equation}
with the boundary condition
\begin{equation} \tilde w(t_2,t_1,\vec r_2-\vec r_1,\vec n_2,\vec n_1) \vert
_{t_2=t_1} = \delta (\vec r_2 -\vec r_1)\delta (\vec n_2 -\vec n_1).
\end{equation}

The term in (1) linear in  $n^{}_0$ is `usual' (incoherent)
contribution to intensity of the electron bremsstrahlung in the
medium derived by summation of the radiation intensities of the
electron interaction with separate atoms of the target.

The term quadratic in   $n_0$ includes  the contribution from the
interference of the brems\-strah\-lung amplitudes on various atoms.
The destructive character of this interference leads to suppression
of the soft radiation intensity, i.e., to the Landau--Pomeranchuk
effect.

For $\omega$ larger than $\omega_{cr}=2I/m^2L$, where $m$ is the
electron mass (for estimation of $\omega_{cr}$ see
\cite{C-2,C-4,C-1.1}), the interference term becomes negligibly
small, and radiation is of pure incoherent character.

For ultrarelativistic particles $(1-v\ll 1)$ it is convenient to
pass in (1) to the small-angle approximation \cite{C-4,C-5}
according to the scheme

$$\vec n_{1,2}=\biggl(1-\frac{\vartheta_{1,2}^{2}}{2}\biggl )\vec n + \vec
\vartheta_{1,2}, \quad d\vec n_{1,2}=d\vec \vartheta_{1,2}~;$$
\begin{equation} f(\vec
n_{1,2})=f(\vartheta_{1,2})=\frac{e}{\pi}\cdot\frac {\vec
\epsilon\vec \vartheta_{1,2}}{\vartheta_{1,2}^{2}+\lambda^2}~,\quad
\lambda = \frac{m}{E} = \gamma^{-1};\end{equation}

$$\sigma_{0}(\vec n_2-\vec n_1)=\sigma_{0}(\vec \vartheta_2-\vec \vartheta_1),\quad
\delta_{0}(\vec n_2-\vec n_1)=\delta_{0}(\vec \vartheta_2-\vec
\vartheta_1),$$

$$\nu_{0}(\vec n_2-\vec n_1)=\nu_{0}(\vec \vartheta_2-\vec \vartheta_1);\quad
w(t_2,t_1,\vec n_2,\vec n_1,\vec k) = w(t_2,t_1,\vec
\vartheta_2,\vec \vartheta_1,\omega) $$

\noindent and further to the Fourier transforms of $f,~\nu ~, w$
$$f(\vec\eta) =  \frac{1}{2\pi}\int \tilde f(\vec\theta)
\exp[i\vec\eta\vec\theta]d\vec\theta = \frac{ie\lambda\vec \epsilon\vec
\eta} {\pi\eta}K_{1}(\lambda\eta);$$
\begin{equation}
\nu(\eta) =  \int \vec
\nu(\vec\theta)e^{i\vec\eta\vec\theta}d\vec\theta =
2\pi\int\sigma_0(\vec \theta)[1-J_0(\eta\theta)]\vec\theta
d\vec\theta;
\end{equation}
$$w(t_2,t_1,\vec \eta_2,\vec \eta_1,\omega) =
\frac{1}{(2\pi)^2}\int \tilde
w(t_2,t_1,\vartheta_2,\vartheta_1,\omega)$$
$$\times\exp[i\vec\eta_2\vec\vartheta_2 - i\vec\eta_1\vec\vartheta_1 ]d\vec\vartheta_1
d\vec\vartheta_2~,$$
where ${\vec\vartheta}_{1(2)}$ are the two-dimensional electron
scattering angles in the plane orthogonal to the electron direction
at the instants of time $t_{1(2)}$, $m$ and $E$ are the electron
mass and its energy, $\vec\theta$ denotes the electron multiple
scattering angle over the interval $t_2-t_1$, $\lambda$ is the
characteristic frequency of the emitted photon, $J_0$ and $K_1$ are
the Bessel and Macdonald functions, respectively.

Consequently, expression (1) is reduced to

$$\frac{dI}{d\omega} = \frac{2\lambda^2 e^2}{\pi^2} \biggl\{ n_0 L
\int K^{2}_{1}(\lambda\eta)\nu(\eta) d\vec\eta   $$
\begin{eqnarray}
-n_0^2\int\limits_{0}^{L}dt_1\int\limits_{0}^{L}dt_2
\int \frac{(\vec \eta_1 \vec \eta_2 )}{\eta_1
\eta_2} K_{1}(\lambda\eta_1)K_{1}(\lambda\eta_2)\nu(\eta_1)\nu(\eta_2)
\end{eqnarray}
$$\times Re [w(t_2,t_1,\vec \eta_2,\vec \eta_1,\omega)]
d\vec\eta_1d\vec\eta_{2}\biggl\},$$
where $w$ satisfies the kinetic equation
$$\frac{\partial w(t_2,t_1,\vec \eta_2,\vec \eta_1,\omega)}{\partial t_2}
=\frac{i\omega}{2}\,(\lambda^2
- \Delta_{\vec \eta_2})w(t_2,t_1,\vec \eta_2,\vec \eta_1,\omega)$$
\begin{equation}
- n_0\nu(\vec \eta_2)w(t_2,t_1,\vec \eta_2,\vec \eta_1,\omega)
\end{equation}
with the boundary condition
\begin{equation}
w(t_2,t_1,\vec \eta_2,\vec \eta_1,\omega) = \delta (\vec \eta_2
-\vec \eta_1).
\end{equation}
The form of equation (7) is similar to the equation for function of
the two-dimensional Schr\"{o}dinger equation with the mass
  $\omega^{-1}$ and the complex potential
\begin{equation}
U(\eta) = - \frac{\omega\lambda^2}{2} - in^{}_0 \nu(\eta)
\end{equation}
and therefore admits of a formal solution in the form of a continual
integral (see, e.g., \cite{C-6}).

\bigskip


{\it 3. Applying the quasiclassical approximation of the Moli\`{e}re
theory to the theory of the LPM effect.} --- The analytical solution
of equation (7) with arbitrary values of $\omega$ is only possible
within the Fokker--Planck approximation is possible:\footnote{
 An explicit expression for $w$ obtained in this approach can be found in \cite{C-5}.}
\begin{equation}
\label{a}\nu(\eta) = a\cdot \eta^{2},
\end{equation}
but at  $\omega = 0$ it is also possible for arbitrary  $\nu
(\eta)$.

In the latter case  $(\omega = 0)$
\begin{equation}
w(t_2,t_1,\vec \eta_2,\vec \eta_1,0) = \delta (\vec \eta_2 -\vec \eta_1)
exp[-n_0\nu(\eta_2)(t_2-t_1)]~,
\end{equation}
and integration over   $t_1,~t_2$ in (6) is carried out trivially,
leading to the simple result
\begin{equation}
\frac{dI(\omega)}{d\omega}\biggl\vert_{\omega=0} = \frac{4\lambda^2
e^2}{\pi} \int K^{2}_{1}(\lambda\eta) (1-\exp[{-n_0}\nu(\eta)L)]\eta d\eta .
\end{equation}

Considering the aforesaid, in the other limiting case $(\omega \gg
\omega_{cr})$ we get
\begin{equation}
\frac{dI(\omega)}{d\omega}\biggl\vert_{\omega \gg \omega_{cr}} = n_0
L\lambda^2 e^2 \int K^{2}_{1} (\lambda\eta)\nu(\eta)\eta d\eta .
\end{equation}

Due to the properties of the Macdonald functions, the main
contribution to the integrals (12), (13) comes from the area
  $0\leq \eta\leq 1/\lambda$. As was shown in classical works of Moli\`{e}re
  \cite{C-7} on the theory of multiple scattering of
  charged particles in a medium, the quantity $\nu(\eta)$ can be represented in this area
  as\footnote{Here the units  $\hbar = c =1$,
$\beta=v/c=1$ are used.}
\begin{equation} \nu(\eta)=4\pi\Bigg(\frac{Z\alpha}{m}\Bigg)^2
\eta^2\,\left[\ln\left(\frac{2}{\eta\,\theta_a^{}}\right)+\frac{1}{2}-C\right],
\end{equation}
where $C=0.57721$ is Euler's constant, and $\theta_a^{}$, refereed
to `a screening angle', depends both on the screening properties of
the atom and on the $\sigma_0(\theta)$-approximation used for its
calculation.

Using the Thomas--Fermi model of  atom \cite{C-8}, Moli\`{e}re
obtained $\theta^{}_a$ for the cases where $\sigma_0$ is calculated
within the Born and quasiclassical approximations:
 \begin{equation}
(\theta^{}_a)_{Born}=1.2\cdot\alpha\cdot Z^{1/3},
\end{equation}
\begin{equation}\label{Mresult}
(\theta^{}_a)_{qcl}=(\theta^{}_a)_{Born} \sqrt{1+3.44\;
(Z\alpha)^2}.
\end{equation}
The latter result is approximate (see critical remarks on its
deviation in \cite{C-9}).

With the technique developed in \cite{C-10}, it is possible to show
\cite{C-11} that for any model of the atom the following relation
holds:
 \begin{equation*}\label{basres} \ln\big
[(\theta^{}_a)_{qcl}\big ]=\ln\big [(\theta^{}_a)_{Born}\big ]+ \Re
\big [\psi (1+iZ\alpha)\big ]+C~ \end{equation*}
or, equivalently,
\begin{equation}\label{Ourresalt}
\ln\big [(\theta^{}_a)_{qcl}\big ]-\ln\big [(\theta^{}_a)_{Born}\big
]=f(Z\alpha)~ ,
\end{equation}
where $\psi$ is a logarithmic derivative of the $\Gamma$-function,
and the Bethe--Maximon function reads
\begin{equation}f(Z\alpha)=(Z\alpha)^2\sum\limits_{n=1}^\infty
\frac{1}{n\left(n^2+(Z\alpha)^2\right)}\ .\end{equation} In the
second order in the parameter $a=Z\alpha$, this result is as follows
\begin{equation}
(\theta^{}_a)_{qcl}=(\theta^{}_a)_{Born} \sqrt{1+2.13\;
(Z\alpha)^2}.
\end{equation}
After the substitution of $\nu(\eta)$ (14) into (13), the
integration is carried out analytically, leading to the following
result:
\begin{equation}
\frac{dI(\omega)}{d\omega}\biggl\vert_{\omega \gg \omega_{cr}} =
\frac{16}{3\pi}\cdot \frac{Z^2\alpha^3}{m^2} \cdot \biggl (\ln
\frac{\lambda}{\theta^{}_a}+ \frac{7}{12}\biggl )\cdot n_0\, L ~.
\end{equation}

Substituting here the numerical values of parameter $\theta^{}_a$
from (16) and (17) corresponding to $Z=79$,
 and introducing the ratio
\begin{equation}
R(\omega)=\frac{\biggl [\frac{dI(\omega)}{d\omega} \biggl
]_{qcl}} {~~\biggl [\frac{dI(\omega)}{d\omega }\biggl
]_{Born}}~~,
\end{equation}
we get $$R(\omega)\vert_{\omega \gg \omega_{cr}}=0.922, $$ which
practically coincides with the normalization factor value $0.93\div
0.94$ introduced in \cite{C-3,C-1.1} for obtaining agreement of the
calculation for $\sigma_0$ in the Born approximation with
experiment.\footnote{Since Migdal used a Gaussian approximation for
the electron distribution, this underestimates the probability of
large angle scatters. These occasional large angle scatters would
produce some suppression for $\omega>\omega_{cr}$, where Migdal
predicts no suppression and where the authors of \cite{C-1.1}
determine the normalization.}

In the other limiting case  $(\omega = 0)$ the performance of
numerical integration in (12), we get the following result for three
thicknesses of the experimental target \cite{C-1.1}:
\begin{equation}
R(\omega)\vert_{\omega=0}=\left\{\begin{array}{ll}0.936,&L=0.001X_0\\
0.961,&L=0.007X_0\;\;\;,\\
0.982,&L=0.060X_0\end{array}\right.
\end{equation}
where $X_0\approx 0.33$ cm is the radiation length of the target
material $(Z=79)$.

It is obvious from  general consideration that when $0 < \omega <
\omega_{cr}$,
\begin{equation}
R(\omega)\vert_{\omega \gg \omega_{cr}}\leq R(\omega) \leq
R(\omega)\vert_{\omega=0}\;\;.  \end{equation}

Finally, let us briefly discuss the accuracy of the Fokker--Planck
approximation that allows an analytical expression to be derived for
$w_0$ and the entire $dI(\omega)/d\omega$ range to be rather simply
calculated (using numerical calculation of triple integrals).

To this end, we will fixe  the parameter  $a$ in expression (10) in
such a way that the results of the exact calculation of
$(dI(\omega)/d\omega)\big\vert_{\omega \gg\omega_{cr}}$
 and its calculation in the Fokker--Planck approximation coincide.
As a result, we get
\begin{equation}
a= 2\pi\biggl (\frac{Z\alpha\sigma}{m} \biggl )^2 \biggl [\ln
\frac{\sigma}{\theta^{}_a}+ \frac{7}{12}\biggl ].
\end{equation}

Now we calculate $(dI(\omega)/d\omega)\big\vert_{\omega = 0}$ using
the relations (22) and (23) and compare the result with the result
obtained using `realistic' (Moli\`{e}re) expression (14) for
$\nu(\eta)$. Then for the ratio
\begin{equation}
\bar R = \frac{\biggl [ \frac{dI(\omega)}{d\omega}\big\vert_{\omega
= 0} \biggl ]_{FP}} {\biggl [
\frac{dI(\omega)}{d\omega}\big\vert_{\omega = 0} \biggl ]_{M}}
\end{equation} we get the following values:
\begin{equation} \bar
R=\left\{\begin{array}{cc}0.947,&L=0.001X_0\\0.890,&L=0.007X_0\\
0.872,&L=0.060X_0\end{array}\right.~.
\end{equation}

It is obvious that the difference (26) from unity is noticeable
higher than the characteristic experimental error. Therefore, it is
clear that Fokker--Planck approximation can be used only for the
qualitative description of the $dI(\omega)/d\omega$ behavior. For
the accurate quantitative analysis it is necessary to use values of
$w$ obtained by the numerical solution of kinetic equation (7). The
results of this analysis together with a detailed comparison with
the experimental data, will be a subject of separate article.

\bigskip


{\it 4. Conclusion.} --- Accounting for the fact that the
calculations for the description of the interaction of electrons
with gold target atoms $(Z \alpha\sim 0.6)$ in \cite{C-3,C-1.1} were
performed using the Born approximation, we managed to show that the
above-mentioned discrepancy between theory and experiment can be
explained both qualitatively and quantitatively if the corrections
to the results of the Born approximation are considered  based on
the quasiclassical approximation of the Moli\`{e}re multiple
scattering theory with the improved value of the screening angular
parameter. From (22) and (23) it follows that the results of the
quasiclassical approach for $dI/d\omega$ cannot be derived from Born
approximation results by multiplying them by the normalization
factor, which is independent of the frequency $\omega$ and target
thickness $L$. However, considering nearly a $3\% $ error of the
experimental data \cite{C-3,C-1.1}, it is clear why multiplication
by the normalization factor helped the authors \cite{C-3,C-1.1} to
get reasonable agreement of the calculations within the Born
approximation with the experimental data.


\vspace{.5cm}

{\small

\begin{thebibliography}{99}

\bibitem{C-2} L.D. Landau and I.Ya. Pomeranchuk, Dokl. Akad. Nauk SSSR,
{\bf 92} 535 (1953); \\E.L. Feinberg and I.Ya. Pomeranchuk, Dokl.
Akad. Nauk SSSR, {\bf 93} 439 (1953); Nuovo Cim. Suppl., {\bf 4} 652
(1956);\\ I.Ya. Pomeranchuk, Dokl. Akad. Nauk SSSR, {\bf 96} 265
(1954), {\bf 96} 481 (1954).

\bibitem{C-4} A.B. Migdal,   Dokl. Akad. Nauk SSSR,  {\bf 96} 49 (1954);
Phys. Rev., {\bf 103} 1811 (1956).

\bibitem{C-4.5} A.B. Migdal, Phys. Rev., {\bf 103} 1811 (1956).

\bibitem{C-5} I.I. Goldman, Zh. Eksp. Teor. Fiz., {\bf 38} 1866 (1960).

\bibitem{Operat1} V.N. Baier and V.M. Katkov, Phys. Rev., {\bf D 57} 3146 (1998).

\bibitem{Operat2}  V.N. Baier, V.M. Katkov, and V.S. Fadin,
Radiation from Relativistic Electrons (in Russian). Atomizdat,
Moscow: Atomizdat, 1973; \\ V.N. Baier, V.M. Katkov, and V.M.
Strakhovenko, Electromagnetic Processes at High Energies in Oriented
Single Crystals. Singapore: World Scientific Publishing Co, 1997.

\bibitem{C-3} B.G. Zakharov, JETP Lett., {\bf 64} 781 (1996).

\bibitem{Z1} B.G.  Zakharov, JETP Lett.,
{\bf 63} 952 (1996); Phys. Atom. Nucl., {\bf 61} 838 (1998).

\bibitem{Crit} B.G. Zakharov, arXiv:hep-ph/9908449 (1999).

\bibitem{Oproverg}  V.N. Baier and V.M. Katkov, arXiv:hep-ph/10202 (1999).

\bibitem{Sh1} N.F. Shul'ga and S.P. Fomin, Probl. Atom. Sci. Technol., {\bf 2} 11
(2003);\\
N.F. Shul'ga, Int. J. Mod. Phys., {\bf A 25} 9 (2010).


\bibitem{showers} T. Stanev et al., Phys. Rev., {\bf D 25} 1291 (1982).

\bibitem{KST} A.H. S{\o}rensen, Z. Phys. {\bf C 53} 595 (1992); \\
B.Z. Kopeliovich, A. Sch\"afer, and A.V. Tarasov, Phys. Rev., {\bf C
59} 1609 (1999).


\bibitem{stars} G. Raffelt and D. Seckel, Phys. Rev. Lett., {\bf 67} 2605 (1991); \\
C.J. Pethick and V. Thorsson, Phys. Rev. Lett., {\bf 72} 1964
(1994).

\bibitem{prev} P.H. Fowler, D.H. Perkins, and K. Pinkau, Philos. Mag., {\bf 4} 1030
(1959); \\E. Lohrmann, Phys. Rev., {\bf 122} 1908 (1961).

\bibitem{C-1.1} P.L. Anthony, R. Becker-Szendy, P.E. Bosted et al., Phys.
Rev.  Lett., {\bf 75} 1949 (1995), {\bf 76} 3550 (1996); Phys. Rev.,
{\bf D 56} 1373 (1997).

\bibitem{C-11} A.V. Tarasov and O.O. Voskresenskaya, ArXiv:1204.3675 [hep-ph].

\bibitem{preprint}  O. Voskresenskaya, A. Sissakian, A.
Tarasov et al., JINR  P2-97-308,  Dubna, 1997 (in Russian).


\bibitem{C-6} R.P. Feynman and A.R. Hibbs, Quantum mechanics and path
integrals, New York: McGrav-Hill, 1965.

\bibitem{C-7} G. Moli\`{e}re, Z. Naturforsch., {\bf 2 a} 133 (1947),
{\bf 3 a} 78 (1948), {\bf 10 a} 177 (1955).

\bibitem{C-8} L.D. Landau and E.M. Lifshitz, Quantum Mechanics,
Moscow: Nauka Publication, 1974.

\bibitem{C-9} W.T. Scott, Rev. Mod. Phys., {\bf 35} 231 (1963).

\bibitem{C-10} O.O. Voskresenskaya, S.R. Gevorkyan, and
A.V. Tarasov, Phys. Atom. Nucl., {\bf 61} 1517 (1998).

\end{thebibliography}

}
\end{document}